\documentclass[lettersize,journal]{IEEEtran}
\usepackage{amsmath,amsfonts}
\usepackage{algorithmic}
\usepackage{algorithm}
\usepackage{array}
\usepackage[caption=false,font=normalsize,labelfont=sf,textfont=sf]{subfig}
\usepackage{textcomp}
\usepackage{stfloats}
\usepackage[hyphens]{url}
\usepackage{hyperref}
\usepackage{verbatim}
\usepackage{graphicx}
\usepackage{cite}
\usepackage{orcidlink}
\begin{document}

\title{A Mixed-Methods Analysis of Repression and Mobilization in Bangladesh's July Revolution Using Machine Learning and Statistical Modeling}

\author{Md. Saiful Bari Siddiqui$^{*}$~\orcidlink{0009-0000-7781-0966}\IEEEmembership{}, 
Anupam Debashis Roy$^{*}$~\orcidlink{0009-0005-0313-6995}.~\IEEEmembership{}
\thanks{$^{*}$These authors contributed equally to this work.

Md. Saiful Bari Siddiqui is a Senior Lecturer at the Department of Computer Science and Engineering, BRAC University, Dhaka, Bangladesh (e-mail: saiful.bari@bracu.ac.bd).

Anupam Debashis Roy is a PhD candidate at the Department of Sociology, University of Oxford, Oxford, United Kingdom (e-mail: anupam.roy@sant.ox.ac.uk).}
\thanks{Manuscript submitted October, 2025.}}





\markboth{July Revolution Analysis, October~2025}%
{Siddiqui \MakeLowercase{\textit{et al.}}: July Revolution Analysis}


\maketitle

\begin{abstract}
The 2024 July Revolution in Bangladesh represents a landmark event in the study of civil resistance: a successful, student-led civilian uprising that overthrew a long-standing authoritarian regime despite facing brutal state repression. This study investigates the central paradox of its success: how state violence, intended to quell dissent, ultimately fueled the movement's victory. We employ a mixed-methods approach. First, we develop a qualitative narrative of the conflict's timeline to generate specific, testable hypotheses. Then, using a disaggregated, event-level dataset, we employ a multi-method quantitative analysis to dissect the complex relationship between repression and mobilisation. We provide a framework to analyse explosive modern uprisings like the July Revolution. Initial pooled regression models highlight the crucial role of protest momentum (measured by a feedback loop effect) in sustaining the movement. To isolate causal effects, we specify a Two-Way Fixed Effects panel model, which provides robust evidence for a direct and statistically significant local suppression backfire effect. Our Vector Autoregression (VAR) analysis provides clear visual evidence of an immediate, nationwide mobilisation in response to increased lethal violence. We further demonstrate that this effect was non-linear. A structural break analysis reveals that the backfire dynamic was statistically insignificant in the conflict's early phase but was triggered by the catalytic moral shock of the first wave of lethal violence, and its visuals circulated around July 16th. A complementary machine learning analysis (XGBoost, out-of-sample R²=0.65) corroborates this from a predictive standpoint, identifying "excessive force against protesters" as the single most dominant predictor of nationwide escalation. We conclude that the July Revolution was driven by a contingent, non-linear backfire, triggered by specific catalytic moral shocks and accelerated by the viral reaction to the visual spectacle of state brutality. 
\end{abstract}


\section{Introduction}
\IEEEPARstart{I}{n} August 2024, the fifteen-year rule of Prime Minister Sheikh Hasina of Bangladesh came to a sudden and dramatic end. After weeks of escalating nationwide protests, she resigned from her post and fled the country. This historic transfer of power was not driven by a military takeover or a traditional political campaign, but by a popular uprising that began in July with a student-led movement against a government job quota system. In a matter of weeks, this initial spark grew into a nationwide fire, as hundreds of thousands of ordinary citizens joined the students, bringing the country to a standstill and achieving a political transformation that had seemed unthinkable just a month earlier.

The July Revolution in Bangladesh is a critical event for understanding modern social movements. In an era where many popular uprisings have struggled to achieve their goals~\cite{OnursalRadicalization2025}~\cite{ChenowethFuture2020}, the Bangladeshi case stands out as a remarkable success. It was a decentralised movement, organised largely by students and powered by ordinary civilians, that managed to overthrow a well-established authoritarian regime. What makes its success even more striking is that it was achieved in the face of extreme state violence. This makes the July Revolution a crucial case study as it provides a powerful model of how a largely non-violent, popular movement can succeed against a heavily armed state, offering vital lessons for scholars and activists alike.

This success presents a central paradox. Authoritarian governments use repression with the clear goal of scaring people off the streets and crushing dissent. The Bangladeshi government followed this exact strategy, resulting in hundreds of deaths and thousands of injuries. Yet, the movement did not die; it grew stronger. This leads to the core question of our research: How and why did the state's most powerful tool of control, repression, fail so completely? This question also presents a significant \textit{methodological challenge.} The revolution's explosive, short-duration nature defies many traditional time-series methods that require long periods of data, demanding a more robust and multi-faceted analytical approach.

To address this challenge, this paper develops and applies a multi-layered quantitative framework designed for analysing short, explosive social movements. Our mixed-methods approach integrates a qualitative historical narrative with a five-stage quantitative analysis, where each stage uses a specific tool to explore different parts of the research problem. This framework moves systematically from broad prediction to sharp causal inference to a nuanced understanding of the conflict's dynamics.

Drawing from a qualitative analysis of the revolution's timeline, we argue that the relationship between repression and mobilisation in Bangladesh was not simple or constant. We developed three core hypotheses. First, we expected that the movement was sustained by its own \textbf{momentum (H1)}. Second, we hypothesised that state repression would, on average, have a \textbf{backfire effect (H2)}, leading to more protest. Finally, and most importantly, we hypothesised that this backfire was \textbf{non-linear and contingent (H3)}, meaning it was not always present but was instead \textit{triggered} by specific, \textbf{catalytic moral shocks}~\cite{MartinBackfire2006} and widely circulated \textbf{visuals} of brutality~\cite{GeiseVisualMobilization2025},~\cite{DoerrVisualAnalysis2013} that fundamentally changed how protesters reacted to state violence.


We test these hypotheses using our layered framework, moving from broad correlational analysis to specific causal and predictive modeling. \textbf{First,} we establish a baseline with a suite of pooled regression models (OLS, Negative Binomial). This initial analysis confirms the powerful effect of protest \textbf{momentum} but reveals the limitations of these models in isolating the true effect of repression due to multicollinearity and omitted variable bias. \textbf{Second,} to address these limitations and move from correlation to causation, we specify a \textbf{Two-Way Fixed Effects panel model.} We use this robust framework first to isolate the direct, local backfire effect and then incorporate a \textbf{structural break analysis} to test our non-linear hypothesis, identifying the precise historical moment the backfire dynamic was triggered. \textbf{Third,} we complement this causal analysis with a \textbf{Vector Autoregression (VAR) model} to visualize the day-to-day dynamic feedback loops between the state and protesters at the national level. \textbf{Finally,} having established the causal story, we employ exploratory machine learning models (XGBoost, Random Forest), validated with a rigorous \textbf{walk-forward cross-validation}, to identify the most powerful \textit{predictive} signals, offering a different lens on the conflict's key triggers.

This paper makes several key contributions:
\begin{itemize}
    \item \textbf{Empirically,} we provide one of the first systematic, data-driven analyses of the landmark 2024 July Revolution.
    \item \textbf{Theoretically,} we refine the ``repression backfire'' literature by providing strong evidence for its non-linear and contingent nature, showing how it can be \textit{triggered} by a \textit{catalytic} event and visuals.
    \item \textbf{Methodologically,} we offer a robust and replicable multi-layered framework for the computational analysis of modern, short-duration, high-intensity social uprisings.
\end{itemize}

\section{Qualitative Analysis and Theoretical Framework}

\subsection{A Brief Chronology of the July Revolution}

The July Revolution was not a single, monolithic event, but a rapid, cascading process that evolved dramatically over a few short weeks. Our qualitative review of news reports, independent media, social media documentation, and human rights reports~\cite{OHCHRProtests2024} identifies three clear phases during the conflict.

\subsubsection{Phase 1: The Quota Reform Protests (Early to Mid-July)}
The conflict's origins were rooted in a long-standing student grievance: a government job quota system that reserved a large percentage of public sector jobs, leaving only 44\% of the total openings for merit-based competition. Protests against this system, under the banner of the ``Student Anti-Discrimination Movement,'' had occurred before, but they gained new traction in early July 2024 following a High Court verdict reintroducing the system. Initially, these demonstrations were classic student protests, largely concentrated on university campuses in the capital city of Dhaka. The demands were specific and policy-focused: a reform of the quota system. While clashes with police and members of the Bangladesh Chhatra League (BCL), a pro-government student wing, were common, the overall intensity was, at first, contained. This period represents the ``baseline'' or low-intensity phase of the conflict, a familiar script of student activism and state containment.

\subsubsection{Phase 2: The Escalation and the Catalytic Moral Shock (Mid-July)}
The dynamics of the conflict shifted irrevocably in mid-July, triggered by a combination of political rhetoric and physical violence. For years, the ruling Awami League party and its affiliate, the Bangladesh Chhatra League (BCL), had employed a political tactic of labeling any form of dissent or opposition with the term ``Rajakar,'' a word evoking the collaborators of the 1971 genocide. This strategy was used to frame political disagreement as a betrayal of the nation's founding principles, effectively stifling criticism. By 2024, this practice had reached a choking point for many citizens, particularly the young generation, leaving little room for legitimate dissent without being branded an anti-nationalist.

It was in this politically charged environment that the initial trigger for escalation occurred. On July 14th, Prime Minister Sheikh Hasina, in a public address, dismissively referred to the student protesters with the phrase ``Rajakar er baccha'' (children of war crimes collaborators). For a movement of young people, many of whom were born long after the war, this was not just an insult but the ultimate expression of the regime's refusal to engage with their grievances. The remark sparked immediate and intense outrage, particularly at Dhaka University, transforming the protest from a single-issue demonstration into a broader stand against the government's entire political culture.

The Prime Minister's comment was followed by a sharp escalation of physical violence. On July 15th, reports and graphic videos began to circulate showing coordinated and brutal attacks on student protesters by BCL activists. The violence intensified dramatically the following day. On July 16th, six protesters, including Abu Sayeed, a 23-year-old student, was shot and killed by police during a protest. His death was not an anonymous statistic; his story and image were shared widely online. The video of his killing was captured on camera, and it almost immediately went viral. The sheer courage of an unarmed Abu Sayeed seemingly greeting bullets with open arms quickly became a powerful, humanising symbol of the state's brutality, marking a critical threshold in the conflict.

The period of intense state violence that followed these events has collectively come to be known as the ``July Massacre.'' In the days immediately following the initial killings, the state's response intensified into a full-scale crackdown. Security forces used live ammunition, rubber bullets, and tear gas indiscriminately. The government attempted to impose an information blackout by shutting down mobile internet services nationwide, and later imposed a shoot-on-sight curfew. The qualitative evidence strongly suggests that this brief, intensely violent period acted as a catalytic moral shock. The conflict was no longer about a specific policy grievance; it became a moral struggle against a government now perceived by many as tyrannical and illegitimate. The visuals of the crackdown, from images of bleeding students to armed BCL members attacking campuses, and the defiance of young people in the face of overwhelming force, created a shared sense of national outrage that fueled the subsequent nationwide uprising.

\subsubsection{Phase 3: The Nationwide Uprising (Late July - August 5th)}
The period following the crackdown was characterized by a dramatic change in the movement's scale, scope, and demands. The state's attempt to intimidate the population had backfired spectacularly. What had been a student-centric protest in Dhaka transformed into a nationwide, popular uprising. A diverse cross-section of society, from university teachers and lawyers~\cite{AlamFear2025} to garment workers and rickshaw pullers, joined the students in the streets, with protests erupting in all eight administrative divisions of the country.

The movement's central demand shifted decisively from ``quota reform'' to the ``one point demand: the immediate resignation of the Prime Minister.'' Protesters openly defied the curfew, with citizens in Dhaka famously coming together at the central Shaheed Minar premises in a coordinated act of civil disobedience. Organizers called for a nationwide ``non-cooperation movement,'' urging immigrant citizens to stop sending remittances, a key contributor to Bangladesh's economy. All these things culminated in the ``Long March to Dhaka,'' where thousands of people from outside the capital began marching towards the city, effectively overwhelming the state's ability to control movement. Faced with a complete loss of authority, Prime Minister Sheikh Hasina resigned and fled on August 5th. This final phase was defined by a new, powerful dynamic where state repression seemed to have lost its deterrent effect entirely, instead functioning as a catalyst for even greater resistance.

\subsection{Theoretical Framework and Literature Review}

Repression, or efforts to suppress either contentious acts or groups and organisations responsible for them, is a predictable response to mobilising efforts in one form or another. When regimes encounter protests that go against their interests, they are likely to respond with repression~\cite{McAdam2001}. Tilly~\cite{Tilly1978} has defined repression as an action by another group that raises the cost of collective action by the contender. Scholars like Titarenko et al.~\cite{Titarenko2001}, studying Belarus in 1990-95 during the country’s transition from authoritarian to hybrid regime, used statistical analysis of police records and argued that repression reduces protests, especially for weak groups, while strong actors adjust their tactics to follow less confrontational tactics. Additionally, Olzak et al.~\cite{Olzak2003} studied the repressive apartheid regime in South Africa through an event-history analysis and found that repression decreased protests in general. 

However, several scholars have argued against this argument, especially when studying democratic or hybrid regimes. Opp et al.~\cite{OppRoehl1990}, studying environmental protests in West Germany through Quantitative empirical analysis using cross-sectional and time-series analysis, argue that repression has no deterrent effect. Also, when analysing the US Civil Rights Movement through case studies, Barkan~\cite{Barkan1984} argued that repression causes outrage and outrage causes further protests. Della Porta~\cite{DellaPorta1995}, studying democratic Italy and Germany over several decades (1960s-1980s), has argued that repression can open political opportunities which lead to even larger protest waves, especially when the repression is viewed as unjust by the protesters. She also argued that micromobisation processes that raise incentives for the protest can also lead to a \textit{nullification of the dampening effect of protests.} Koopmans~\cite{Koopmans1997} did a quantitative empirical analysis using cross-sectional and time-series analysis of the extreme right-wing mobilization in Germany (1990–1994) and concluded that situational repression (police intervention) tended to escalate protests, particularly when repression was inconsistent, while institutional repression (bans, trials) had a deterrent effect, reducing participation in extreme-right mobilisation. 

In a more comprehensive study, Carey~\cite{Carey2006} analyses six Latin American and three African countries over approximately 12 years whose regimes are characterised as democratic, autocratic, and semi-democratic and finds that there is a reciprocal relationship between protest and repression. In most cases, protest leads to repression, and repression leads to protest. This suggests a feedback loop where hostile actions by one actor (government or population) provoke hostile actions by the other. She also finds that both \textit{protest and repression are autoregressive}, meaning that once protest or repression starts, it tends to continue. Protest at time $t$ led to protest at time $t+1$ in 17 out of 18 of her cases, and repression at time $t$ led to repression at time $t+1$ in 14 out of 18 cases. Biggs~\cite{Biggs2003} has a similar finding where he theorised that positive feedback takes place when protests in time $t$ leads to an increase in protests in time $t+1$, meaning protests incite further protests through a process of emulation. This happens because, in these situations, the motivation to participate in protests increases with the number of participants as the expected collective benefits increase and costs decrease, while the moral obligation to participate also increases~\cite{Biggs2003}.

Additionally, scholars like Hess and Martin~\cite{MartinBackfire2006} have found a greater incidence of protests after particular repressive actions. They use terms like backfire and transformative events to explain why some events are more likely than others to create further protests. Hess and Martin~\cite{MartinBackfire2006} define transformative events as crucial turning points for a social movement that dramatically increase or decrease the rates of mobilisation. Tilly~\cite{Tilly1978} supports this idea that repression by authorities has the potential of becoming a transformative event by causing a significant increase in the cost of mobilising and organising work, whereas McAdam et al.~\cite{McAdam2001} support the idea that transformative events can lead to greater mobilisation. Hess and Martin~\cite{MartinBackfire2006} further argue that repressive events lead to “backfire”, which is defined as a “public reaction of outrage to an event that is publicised and perceived as unjust. 

Backfire has been defined in the literature as a phenomenon that sees unjust acts, such as violent repression, recoils against those who perpetrate it, and leads to a power shift by increasing the internal solidarity of resistance campaigns by increasing external support for the resistance campaigns, and decreasing support for the perpetrators~\cite{BjorkJames2024}. This type of action by the government leads to a tarnishing of their reputation and, if the movement cannot be totally suppressed by these events, then also their weaknesses get publicly revealed~\cite{BjorkJames2024}. 

To answer the question of how these backfires come about, scholars have built the frame of \textit{transformative events}, which are “short, contingent conjunctures in which agency collides with structure to rapidly reconfigure political trajectories” that “disrupt taken-for-granted routines, create sharp uncertainty, and can open or foreclose paths”~\cite{BosiDavis2017, Davis2015}. In repressive contexts, particular events, especially \textit{illegitimate, visibly unjust repression}, produce \textit{moral shock} that (1) delegitimizes incumbents, (2) diffuses anger and solidarity rapidly, (3) lowers coordination barriers, and (4) can flip state violence into regime-weakening \textit{backfire}~\cite{MartinBackfire2006, Shultziner2018}.

A “transformative shock” frame, then, names those critical events whose \textit{moral} transgression becomes public enough to expose a regime as \textit{indefensible} to its own population, thereby catalyzing path-changing sequences: mobilization surges, elite splits, and re-alignment of opportunity/threat perceptions. This frame both extends the concept of “political shocks”~\cite{Rohlinger2009} and sharpens the idea of “transformative events”~\cite{BosiDavis2017, Davis2015} by foregrounding the \textit{ethical breach} (moral shock) as the distinguishing ignition mechanism—and by recognizing that these events can also \textit{foreclose} peaceful trajectories~\cite{BosiDavis2017, Davis2015}.

The concept of transformative events has been more prominent in recent work on social movement and contentious politics. Scholars find some short-term and proximate events that alter political trajectories and reform the structure-agency dynamic~\cite{BosiDavis2017, Davis2015}. However, the terminology of "transformative event” may seem tautological. Events are too easily understood as transformative because they are subsequently tagged as having altered politics. As Davis~\cite{Davis2015} notices in his account of the 1916 Easter Rising, "only really become 'events' after the fact," because it is then proclaimed a turning point. Bosi and Davis~\cite{BosiDavis2017} also recognize the slip between "what causes a particular event" and "what makes it transformative" and mention the risk of circularity. At its worst, the idea amounts to an ex post facto fallacy: one calls something important only after the fact due to outcomes that ensued and not according to processes that would have been apparent then.

Therefore, it seems more logical to use a different formulation to understand similar political events. In this paper, we call the key rupture of killing students (along with the circulated visuals of them) by the police as \textit{catalytic moral shocks}. This is not a new concept, but rather one based on existing terminologies used in social movement studies. Shultziner~\cite{Shultziner2018}, for example, highlights the role of \textit{moral shocks}, publicly perceived violations of ethical norms, that can catalyse mobilisation by delegitimizing regimes and lowering the costs of defiance. In this formulation, the outcome is not assumed, but contingent on the interaction of actors, resources, and constraints~\cite{Collier1991, Pierson2004}. This framing avoids tautology by embedding events in broader temporal processes. Likewise, Sewell’s \cite{Sewell1996} notion of “eventful temporality” emphasises that events matter not in themselves but because they reorder social structures through mechanisms of interpretation, mobilisation, and institutional change. Building on this insight, we propose the concept of \textbf{catalytic moral shocks}. 

Unlike the ambiguous label “transformative event,” this term specifies both mechanism and distinguishing feature. Such shocks are \textit{catalytic} because they accelerate or redirect political processes without predetermining outcomes, and they are \textit{moral} because what sets them apart is the public recognition of a regime’s actions as ethically indefensible. This recognition creates delegitimation, diffuses outrage, and lowers barriers to collective action. Not all catalytic events transform political trajectories, but those that do can be traced through identifiable mechanisms: mobilisation surges, shifts in elite alignments, and institutional reconfigurations. For example, repression that generates moral shock may produce backfire~\cite{MartinBackfire2006}, undermine the authority of incumbents~\cite{Shultziner2018}, or foreclose compromise solutions~\cite{BosiDavis2017}.

Therefore, to avoid the tautology of labeling an event ``transformative'' \textit{a priori}, we propose a more precise formulation. We define the key rupture of mid-July---the Prime Minister's inflammatory rhetoric, the coordinated BCL attacks, and the widely circulated visuals of Abu Sayeed's killing---as a potential \textbf{catalytic moral shock}. This term specifies both the mechanism and the distinguishing feature of the event. Such a shock is \textit{catalytic} because it can rapidly accelerate or redirect political processes without predetermining a single outcome, and it is \textit{moral} because what sets it apart is the public recognition of the regime's actions as an ethically indefensible breach of norms.

This framework allows us to make a clear, testable claim. We argue that this catalytic moral shock fundamentally altered the conflict's dynamics, unlocking a powerful backfire effect and turning protest momentum into a revolutionary cascade. We can then use our quantitative analysis to test this proposition directly. By proving, \textit{ex post facto}, that this shock did indeed produce a structural break in the repression-mobilization dynamic, we can then conclude that it was, in fact, a transformative event. We argue that because of the moral shock, the trajectory of the movement was positively changed, a feedback loop was strengthened (which we are calling \textit{momentum} in this paper), and backfire went into effect, turning the moral shock into a transformative event. This approach allows us to separate the cause from the effect, providing a logically sound foundation for our analysis.


\subsection{Hypothesis Development}
This qualitative narrative of a movement sustained by its own energy and catalysed by a violent shock leads us to three core hypotheses, which form the basis of our quantitative analysis.

First, the narrative of protest waves growing day by day, even before the crackdown, points to the importance of internal momentum~\cite{LiljaResistance2017}~\cite{GonzalezProtestDiffusion2025}. A successful protest on one day signals to others that the movement is viable and lowers the perceived risk of participation on the next. This leads to our first hypothesis:

\vspace{0.5em} 
\noindent
\textbf{H1 (Momentum):} \textit{The level of protest mobilisation on a given day will be positively and significantly associated with the level of mobilisation on the preceding day.}
\vspace{0.5em}

Second, the dramatic escalation of the movement immediately following the start of the escalated state repression, particularly Abu Sayeed's death and its heart-touching visuals, aligns with the well-established ``suppression backfire'' theory~\cite{SimpsonViolent2018},~\cite{HorizonsBackfire2023},~\cite{OBrienChina2014}. According to this theory, state violence, intended to deter, can instead increase outrage, legitimise the protesters' cause, and draw in new participants who were previously on the sidelines. This leads to our second hypothesis:

\vspace{0.5em}
\noindent
\textbf{H2 (Backfire):} \textit{The level of state repression, particularly lethal violence accompanied with visuals, will be positively and significantly associated with the level of protest mobilization on subsequent days.}
\vspace{0.5em}

\begin{figure*}[t!]
    \centering
    \includegraphics[width=0.48\textwidth]{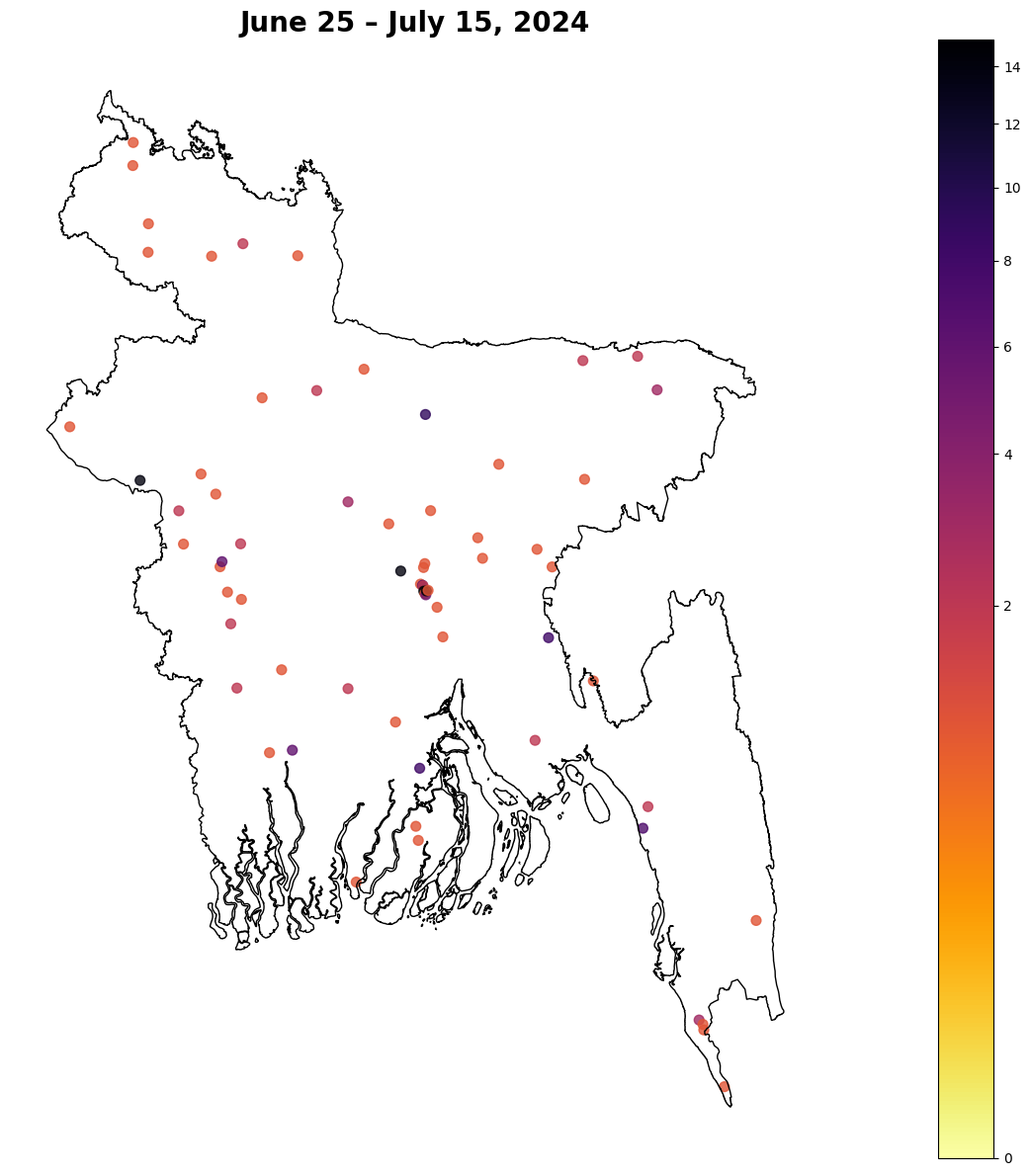} 
    \hfill
    \includegraphics[width=0.48\textwidth]{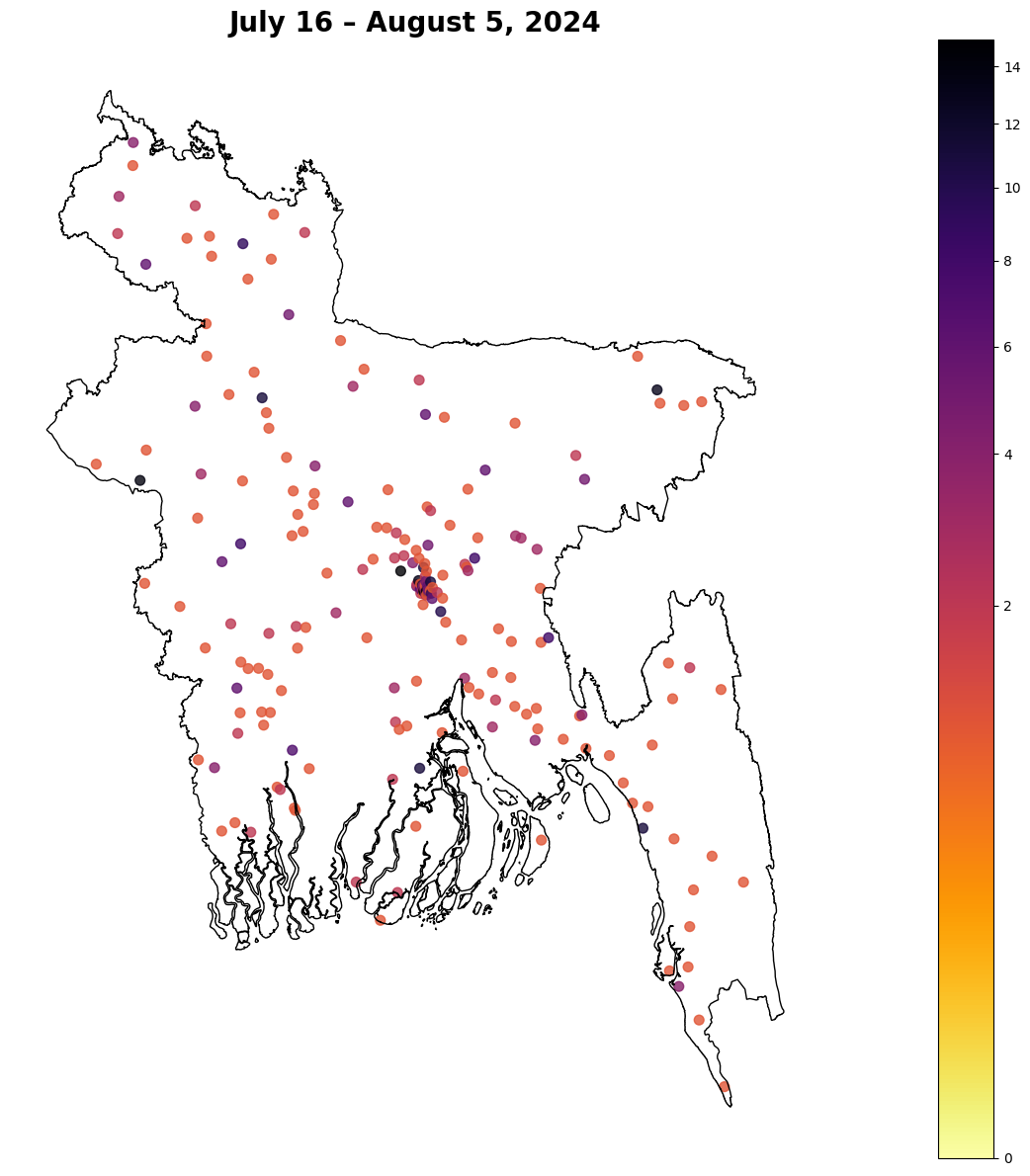}
    \caption{Geographic Distribution of Protest Events Before and After the July 16th Turning Point. The map on the left shows protest events from June 25 - July 15, concentrated primarily in the Dhaka division. The map on the right shows the nationwide explosion of protest events from July 16 - August 5.}
    \label{fig:maps}
\end{figure*}

Finally, and most critically, our qualitative analysis strongly suggests that the backfire effect was not a constant, linear feature of the conflict. The dynamics appeared to be fundamentally different before and after the mid-July crackdown. The violence of July 15-16th, symbolized by the death of Abu Sayeed, appears to have acted as a ``tipping point'' or a \textit{trigger} that \textit{unlocked} a much stronger and more widespread popular response to repression, as we can see from the nationwide explosion of protest events after July 16th from Figure \ref{fig:maps}. This suggests a contingent relationship, leading to our third and central hypothesis:

\vspace{0.5em}
\noindent
\textbf{H3 (Non-Linearity / Structural Break):} \textit{The positive relationship between repression and mobilisation will be contingent on a catalytic event (moral shock). We hypothesize that the backfire effect will be statistically insignificant or weak in the period before July 16th, but positive and significantly stronger in the period on and after this date.}
\vspace{1em} 

These three hypotheses, grounded in the historical narrative of the July Revolution, provide a clear roadmap for our quantitative investigation. The next section details the data and the multi-layered methodological framework we employ to systematically test them.

\section{Data and Quantitative Methodology}

To empirically test our hypotheses, we constructed a novel event dataset and developed a multi-layered quantitative framework. This section details our research design. First, we describe the primary data source and the variables used to measure our core concepts. Second, we outline the data preparation and feature engineering process. Third, we explain the econometric models chosen for causal inference, including the Two-Way Fixed Effects and Vector Autoregression specifications. Finally, we detail the machine learning models used for our exploratory predictive analysis.

\subsection{Dataset}

Our analysis relies on event data from the Armed Conflict Location \& Event Data Project (ACLED), a comprehensive, publicly available database that tracks political violence and protest events worldwide \cite{acled_source}. ACLED is uniquely suited for this study due to its high temporal frequency (daily), spatial precision (georeferenced), and disaggregated event typology, allowing for a granular analysis of the conflict's day-to-day dynamics. Data is systematically coded by ACLED from a curated collection of open-source materials, including international and local news reports, social media, and reports from non-governmental organizations. Our primary dataset covers all reported events in Bangladesh from June 1, 2024, to August 5, 2024. This timeframe was chosen to cover the entire timeline of the movement, starting from the day of the High Court verdict and concluding on the day the regime collapsed.

From the raw ACLED data, we constructed several variables to operationalise our core concepts. Our primary dependent variable, \textbf{Mobilisation}, is measured as `\texttt{total\_events}`, a daily count of all protest events, which is an aggregate of ACLED's `protests`, `riots`, and `violent demonstrations` sub-types. Our primary independent variable, \textbf{Repression}, is measured using several indicators. Lethal repression is measured as `\texttt{total\_fatalities}`, a daily count of all deaths resulting from political violence. Non-lethal and visual repression are captured by disaggregated ACLED event types, most notably `Arrests` and `Excessive force against protesters`.

\subsection{Data Preparation and Feature Engineering}

The raw, event-level ACLED data was processed into two distinct structured formats for our analysis: (1) a \textbf{national-level daily time-series}, created by aggregating all events across the country for each day; and (2) a \textbf{panel dataset}, structured at the \textbf{division-day} level, with observations for each of the eight administrative divisions of Bangladesh for each day in our sample period.

To prepare the data for modelling, we engaged in a multi-step feature engineering process. First, to mitigate the risk of simultaneity bias and to model the reactive nature of protest, all predictor variables were lagged. Our exploratory analysis, which compared the predictive performance of models with 1-day, 2-day, and 3-day lag structures, found that a \textbf{2-day lag structure} consistently provided the highest out-of-sample predictive accuracy. This finding, which we attribute to the information delays caused by state-imposed internet shutdowns, informed the specification of our final predictive models.

Second, we created several new columns to capture specific theoretical concepts. A series of binary dummy variables was created to represent key institutional actions by the state, such as `Curfew` and `Mobile Internet Shutdown`. We also created a `Major Event` dummy variable, manually coded to equal `1` on days identified in our qualitative narrative as having a major historical significance (e.g., the death of Abu Sayeed, the ``Long March to Dhaka''), allowing us to control for these unique moments. All feature engineering was conducted prior to the splitting of data for cross-validation to prevent data leakage.

\subsection{Statistical Modelling for Causal Inference}

To test our primary hypotheses regarding the causal effect of repression, we employ a series of econometric models, each chosen for its ability to address specific statistical challenges.

\subsubsection{Pooled Regression Models (OLS and Negative Binomial)}
We begin our analysis with a suite of pooled regression models to establish baseline correlations and to serve as a benchmark against which our more complex models can be compared. Although our dependent variable is a count of protest events, we first specify an Ordinary Least Squares (OLS) model~\cite{AsteriouAppliedEcon2023}. This choice is deliberate; OLS is a robust and easily interpretable linear benchmark, and by applying a square-root transformation to the dependent and independent variables, we can stabilise the variance and mitigate the influence of extreme outliers that are characteristic of conflict data. This provides a valuable first approximation of the relationships in our data. The OLS model is specified as:
\begin{equation}
\begin{split}
    \sqrt{\text{Events}_{it}} = \beta_0 &+ \beta_1\sqrt{\text{LocalFatal}_{it-1}} + \beta_2\sqrt{\text{ElseFatal}_{it}} \\
                                       &+ \beta_3\sqrt{\text{LocalEvents}_{it-1}} + \beta_4\sqrt{\text{ElseEvents}_{it}} \\
                                       &+ \beta_5\text{Days}_{t} + \beta_6\text{Pop}_{i} + \epsilon_{it}
\end{split}
\end{equation}
where $i$ denotes the division and $t$ denotes the day. The coefficients ($\beta$) represent the linear relationship between the transformed predictors and the transformed outcome.

While the OLS model provides a useful linear approximation, another robust approach is to model the event counts directly. Given that our data exhibits significant overdispersion (i.e., the variance is much larger than the mean), a standard Poisson model is inappropriate. We therefore specify a \textbf{Negative Binomial (NB) model}, which is the gold standard for overdispersed count~\cite{HilbeNB2011}. The NB model assumes that the expected count of events, $\mu$, follows a log-linear relationship with the predictors. It also includes a dispersion parameter, $\alpha$, to account for the excess variance. The model is fully specified by its conditional mean:
\begin{equation}
\begin{split}
    \ln(\mu_{it}) = \beta_0 &+ \beta_1\sqrt{\text{LocalFatal}_{it-1}} + \beta_2\sqrt{\text{ElseFatal}_{it}} \\
                           &+ \beta_3\sqrt{\text{LocalEvents}_{it-1}} + \beta_4\sqrt{\text{ElseEvents}_{it}} \\
                           &+ \beta_5\text{Days}_{t} + \beta_6\text{Pop}_{i}
\end{split}
\end{equation}
where the variance is modeled as $\text{Var}(Y_{it}) = \mu_{it} + \alpha\mu_{it}^2$. The coefficients ($\beta$) from the NB model are interpreted as Incidence Rate Ratios (IRR) after exponentiation ($e^\beta$), representing the multiplicative change in the expected count of events for a one-unit change in a predictor. The significance of the dispersion parameter, $\alpha$, serves as a formal statistical test for overdispersion. While these pooled models are useful for understanding broad associations and the importance of protest momentum, they are vulnerable to omitted variable bias as they do not account for the panel structure of the data, a limitation we address with our next specification.

\subsubsection{Two-Way Fixed Effects (TWFE) Panel Model}
To move from correlation to a more robust causal estimate of the backfire effect, our core causal model is a Two-Way Fixed Effects (TWFE) panel regression~\cite{BaltagiPanel2013}. This specification is crucial for addressing the omitted variable bias that affects pooled models. The general form of the model we initially specify includes our key time-varying predictors:
\begin{equation}
\begin{split}
    \sqrt{\text{Events}_{it}} = & \beta_1\sqrt{\text{LocalFatal}_{it-1}} + \beta_2\sqrt{\text{LocalEvents}_{it-1}} \\
                               & + \beta_3\sqrt{\text{ElseFatal}_{it}} + \beta_4\sqrt{\text{ElseEvents}_{it}} \\
                               & + \beta_5\text{Days}_{t} + \beta_6\text{Pop}_{i} \\
                               & + \alpha_i + \gamma_t + \epsilon_{it}
\end{split}
\end{equation}
where $Y_{it}$ represents the mobilization level (square-root transformed) in division $i$ on day $t$. The term $\alpha_i$ represents a full set of division-specific fixed effects, and $\gamma_t$ represents a full set of day-specific fixed effects.

The power of the TWFE specification lies in its ability to control for unobserved confounding variables. The division fixed effects ($\alpha_i$) absorb the influence of all stable, time-invariant characteristics of each region, such as its population, political culture, or economic status. Any variable that is constant for a given division over the sample period becomes perfectly collinear with that division's fixed effect and is therefore dropped from the model. Similarly, the time fixed effects ($\gamma_t$) absorb the influence of all national-level shocks or trends that affect all divisions on a given day, such as a major political announcement or a nationwide internet shutdown.

As a result of this process, our fully specified model necessarily drops the time-invariant (`Population`) and nation-constant (`Elsewhere events`, `Elsewhere Fatalities`, etc.) predictors. The final, estimable model isolates the coefficients on the purely local, time-varying predictors:

\begin{equation}
\begin{split}
    \sqrt{\text{Events}_{it}} = & \beta_1\sqrt{\text{LocalFatal}_{it-1}} + \beta_2\sqrt{\text{LocalEvents}_{it-1}} \\
                               & + \alpha_i + \gamma_t + \epsilon_{it}
\end{split}
\end{equation}

The coefficient of primary interest, $\beta_1$, therefore represents the estimated effect of a change in local repression on subsequent local mobilization, having controlled for both the unique characteristics of each division and any shared national-level shocks. 

To test our non-linear hypothesis (H3) that this effect was contingent on a catalytic event, we augment this final specification with a structural break analysis~\cite{BaiPerron2003}. We introduce a dummy variable, $D_t$, which equals 0 for the period before a specific cutoff date (e.g., July 16th) and 1 for the period on and after that date. We then interact this dummy with our repression variable. The augmented model takes the form:
\begin{equation}
\begin{split}
    \sqrt{\text{Events}_{it}} = & \beta_1\sqrt{\text{LocalFatal}_{it-1}} \\
                               & + \beta_2(\sqrt{\text{LocalFatal}_{it-1}} \times D_t) \\
                               & + \beta_3\sqrt{\text{LocalEvents}_{it-1}} + \alpha_i + \gamma_t + \epsilon_{it}
\end{split}
\end{equation}
In this specification, the coefficient $\beta_1$ now represents the effect of repression \textit{before} the catalytic moral shock. The coefficient on the interaction term, $\beta_2$, represents the \textit{change} in the effect of repression \textit{after} the catalytic moral shock. A statistically significant $\beta_2$ provides a formal test for a structural break in the repression-mobilization dynamic.

\subsubsection{Vector Autoregression (VAR)}
While our Two-Way Fixed Effects model provides a robust estimate of the \textit{causal effect} of local repression, it is an essentially static model that estimates an average effect over time. To complement this analysis and to explicitly model the \textit{dynamic, temporal interplay} between repression and mobilization, we specify a Vector Autoregression (VAR) model on our national-level daily time-series. A VAR model is uniquely suited for capturing the complex, multi-day feedback loops inherent in a conflict situation~\cite{PedroniSPVAR2013},~\cite{CanovaPVAR2013}. It can capture not only how repression affects future mobilisation, but also how mobilisation affects future repression.

A VAR model is a system of multivariate time-series equations where each variable is regressed on its own lagged values and the lagged values of all other variables in the system. A critical prerequisite for a stable VAR model is that all time series are stationary. We test both our `total fatalities` and `mobilisation events` series for stationarity using the Augmented Dickey-Fuller (ADF) test. As both series were found to be non-stationary in their levels, we apply a first-difference transformation to both, creating two new stationary series: $\Delta \text{Fatalities}_t$ and $\Delta \text{Mobilization}_t$. The optimal lag length, $p$, for the model is determined by minimizing the Akaike Information Criterion (AIC). For a 2-variable VAR(p) model, the system of equations is:
\begin{equation}
\begin{split}
    \Delta \text{Fatalities}_t = c_1 + \sum_{i=1}^{p} \alpha_{1i} \Delta \text{Fatalities}_{t-i} \\
                               + \sum_{i=1}^{p} \beta_{1i} \Delta \text{Mobilization}_{t-i} + u_{1t}
\end{split}
\end{equation}
\begin{equation}
\begin{split}
    \Delta \text{Mobilization}_t = c_2 + \sum_{i=1}^{p} \alpha_{2i} \Delta \text{Fatalities}_{t-i} \\
                                 + \sum_{i=1}^{p} \beta_{2i} \Delta \text{Mobilization}_{t-i} + u_{2t}
\end{split}
\end{equation}
where $c_1$ and $c_2$ are intercepts, $\alpha$ and $\beta$ are the coefficient matrices to be estimated, and $u_{1t}$ and $u_{2t}$ are white-noise error terms.

The primary output for our analysis is the \textbf{Impulse Response Function (IRF)}. The individual coefficients in a VAR model are difficult to interpret directly. An IRF, however, provides a clear and intuitive visualization of the system's dynamics. It traces the dynamic effect of a one-standard-deviation ``shock'' (an unexpected increase) in one variable on the future values of another variable over a specified time horizon. By simulating a shock to $\Delta \text{Fatalities}_t$ and observing the response of $\Delta \text{Mobilization}_t$ over the subsequent days, the IRF provides a powerful visual representation of the suppression backfire's timing, magnitude, and persistence at the national level.

\subsection{Predictive Modelling with Machine Learning}
While our econometric models are specified to test precise causal hypotheses, they are necessarily parsimonious and rely on strong theoretical assumptions. To provide a different, complementary, and more exploratory lens on the conflict, we employ a predictive modeling approach using machine learning. The goal of this stage is not causal inference in the traditional sense, but rather to answer a different and equally important question: given the full, high-dimensional informational landscape of the conflict, which signals had the highest predictive utility for forecasting the next wave of mobilization? By identifying the most important predictors without the constraints of a linear functional form, we can uncover empirical regularities and potential mechanisms that might be missed by more traditional models, such as complex interaction effects or non-linear thresholds.

\subsubsection{XGBoost and Random Forest}
For this task, we utilise two powerful, tree-based ensemble models: \textbf{Random Forest} and \textbf{XGBoost} (Extreme Gradient Boosting)~\cite{ChenXGBoost2016}. These models are particularly well-suited for this analysis~\cite{NichoAXGBoost2025} for several reasons. First, they are inherently non-linear, capable of capturing complex relationships in the data without requiring prior specification. Second, they are robust to the inclusion of a large number of predictors and are not as sensitive to multicollinearity as regression models.

A \textbf{Random Forest} is a bagging algorithm that builds a large ensemble of decorrelated decision trees~\cite{BreimanRF2001}. For a set of $B$ trees, the final prediction for a regression problem is the average of the predictions from each individual tree:
\begin{equation}
    \hat{f}_{\text{rf}}(x) = \frac{1}{B} \sum_{b=1}^{B} T_b(x)
\end{equation}
where $T_b(x)$ is the prediction of the $b$-th tree. The decorrelation is achieved by training each tree on a bootstrapped sample of the data and, crucially, by only considering a random subset of features at each split.

\textbf{XGBoost} is a boosting algorithm that builds trees sequentially to form an additive model. The prediction is the sum of the predictions of all trees, where each new tree is trained to predict the residual errors of the preceding ones:
\begin{equation}
    \hat{y}_i = \sum_{k=1}^{K} f_k(x_i), \quad f_k \in \mathcal{F}
\end{equation}
where $K$ is the number of trees and $f_k$ is the prediction of the $k$-th tree. The algorithm minimizes a regularized objective function to prevent overfitting and control for model complexity.

The primary interpretive tool for these models is their \textbf{feature importance} ranking. This metric, typically calculated as the total gain or the total number of times a feature is used for a split across all trees, quantifies how much each variable contributed to the model's predictive accuracy.

\subsubsection{Model Validation: Walk-Forward Cross-Validation}
A critical challenge in any predictive modeling exercise with time-series data is the risk of overfitting and data leakage from the future. Standard k-fold cross-validation is inappropriate as it would allow the model to be trained on future events to ``predict'' the past. To ensure an honest, out-of-sample evaluation of our models' true forecasting ability, we employ a rigorous \textbf{walk-forward cross-validation} procedure~\cite{HyndmanForecasting2021}.

This procedure rigorously simulates a real-world forecasting scenario. For each day $t$ in our validation set, the model is trained on all available historical data from day 1 up to day $t-1$. It then makes a single prediction for the outcome on day $t$. This out-of-sample prediction is stored, the window of training data is expanded to include day $t$, and the process is repeated to predict day $t+1$. The final performance metrics (e.g., R-squared, Mean Absolute Error) are then calculated on the complete set of these out-of-sample predictions, providing a robust and realistic measure of the models' true predictive power.
This procedure is formally described in Algorithm \ref{alg:walk_forward_cv}.

\begin{algorithm}[H]
\caption{Walk-Forward Cross-Validation for Time-Series Models.}\label{alg:walk_forward_cv}
\begin{algorithmic}[1]
\STATE \textbf{Input:} 
\STATE \hspace{0.5cm} Feature matrix $\mathbf{X}$ of size ($N \times M$) sorted by time
\STATE \hspace{0.5cm} Target vector $\mathbf{y}$ of size ($N \times 1$) sorted by time
\STATE \hspace{0.5cm} Minimum training size $N_{min}$
\STATE \hspace{0.5cm} Model class $\mathcal{M}$ (e.g., XGBoost, Random Forest)
\STATE \textbf{Initialize:}
\STATE \hspace{0.5cm} Empty list of true values, $\mathbf{y}_{true}$
\STATE \hspace{0.5cm} Empty list of predicted values, $\mathbf{y}_{pred}$
\STATE 
\STATE \textbf{Procedure:} {\textsc{WalkForwardCV}}$(\mathbf{X}, \mathbf{y}, N_{min}, \mathcal{M})$
\STATE \FOR{$t = N_{min}$ \TO $N-1$}
\STATE \hspace{0.5cm} // Define the expanding training window
\STATE \hspace{0.5cm} $\mathbf{X}_{train} \gets \mathbf{X}[0:t, :]$
\STATE \hspace{0.5cm} $\mathbf{y}_{train} \gets \mathbf{y}[0:t]$
\STATE 
\STATE \hspace{0.5cm} // Define the single-step-ahead test window
\STATE \hspace{0.5cm} $\mathbf{x}_{test} \gets \mathbf{X}[t, :]$
\STATE \hspace{0.5cm} $y_{test} \gets \mathbf{y}[t]$
\STATE 
\STATE \hspace{0.5cm} // Train the model on all available past data
\STATE \hspace{0.5cm} $model \gets \mathcal{M}()$
\STATE \hspace{0.5cm} $model.\textsc{fit}(\mathbf{X}_{train}, \mathbf{y}_{train})$
\STATE 
\STATE \hspace{0.5cm} // Predict the single next time step
\STATE \hspace{0.5cm} $\hat{y}_{pred} \gets model.\textsc{predict}(\mathbf{x}_{test})$
\STATE 
\STATE \hspace{0.5cm} // Store the prediction and the true value
\STATE \hspace{0.5cm} Append $y_{test}$ to $\mathbf{y}_{true}$
\STATE \hspace{0.5cm} Append $\hat{y}_{pred}$ to $\mathbf{y}_{pred}$
\STATE \ENDFOR
\STATE \textbf{return} $\mathbf{y}_{true}, \mathbf{y}_{pred}$
\end{algorithmic}
\end{algorithm}

\section{Quantitative Results \& Discussion}

This section presents the results of our four-stage quantitative analysis. We proceed in the order outlined in our methodological framework, moving to exploratory predictive modeling from our core causal and dynamic analyses.

\subsection{Baseline Analysis: The Role of Momentum}

We begin our quantitative analysis by establishing baseline correlations with a set of pooled regression models. This initial step serves two purposes: first, to test our hypothesis regarding protest momentum powered by a feedback loop (H1); and second, to assess the broad relationships in the data before applying more stringent causal controls. We specified two distinct models, presented in Table \ref{tab:pooled_models}. The first is an Ordinary Least Squares (OLS) model on square-root-transformed variables, which serves as a robust linear benchmark. The second is a Negative Binomial (NB) model, suited to the count-based, overdispersed nature of our protest data. A diagnostic test for multicollinearity revealed no cause for major concern among the predictors (all VIF < 5), allowing us to proceed with interpreting these initial models.

\begin{table}[h!]
\centering
\caption{Pooled Regression Models Predicting Protest Mobilization}
\label{tab:pooled_models}
\begin{tabular}{lcc}
\hline \hline
 & \textbf{(1) OLS} & \textbf{(2) Negative Binomial} \\
 & $\sqrt{\text{Events}}$ ($\beta$) & Events (IRR) \\
\hline
\textit{Local Dynamics} & & \\
$\sqrt{\text{LocalFatal}_{t-1}}$ & 0.157$^{***}$ & 1.052 \\
 & (0.047) & (0.959 -- 1.154) \\
$\sqrt{\text{LocalEvents}_{t-1}}$ & 0.199$^{***}$ & 1.289$^{***}$ \\
 & (0.049) & (1.119 -- 1.484) \\
 & & \\
\textit{National Dynamics} & & \\
$\sqrt{\text{ElseFatal}_{t}}$ & -0.036$^{*}$ & 0.903$^{***}$ \\
 & (0.021) & (0.849 -- 0.960) \\
$\sqrt{\text{ElseEvents}_{t}}$ & 0.284$^{***}$ & 1.578$^{***}$ \\
 & (0.031) & (1.419 -- 1.755) \\
 & & \\
\textit{Controls} & & \\
Days Since Start & 0.002 & 1.002 \\
 & (0.002) & (0.995 -- 1.010) \\
Population (Millions) & 0.0369$^{***}$ & 1.062$^{***}$ \\
 & (0.003) & (1.051 -- 1.073) \\
\hline
Dispersion ($\alpha$) & -- & 0.232$^{***}$ \\
 & & (0.054) \\
\hline
Observations & 488 & 488 \\
Adj. $R^2$ / Pseudo $R^2$ & 0.587 & 0.245 \\
\hline \hline
\multicolumn{3}{l}{\footnotesize Standard errors in parentheses for OLS; 95\% CI for NB IRR.} \\
\multicolumn{3}{l}{\footnotesize $^{*}p<0.1$; $^{**}p<0.05$; $^{***}p<0.01$} \\
\end{tabular}
\end{table}

Despite their different assumptions, both models converge on a single, powerful conclusion: protest momentum was a dominant feature of the July Revolution. In the OLS model (Column 1), the coefficients for prior-day local events ($\beta=0.199$, $p < 0.001$) and events elsewhere ($\beta=0.284$, $p < 0.001$) are both positive and highly significant. The Negative Binomial model (Column 2) confirms this and allows for a more direct interpretation; a one-unit increase in the square root of prior-day local events is associated with a 29\% increase in the expected number of current-day events (IRR = 1.29), while a similar increase in events elsewhere is associated with a powerful 58\% increase (IRR = 1.58). This remarkable consistency provides strong, robust support for our first hypothesis (H1), establishing that the revolution was a highly dynamic process, sustained and amplified by prior mobilization.

However, when examining the direct effect of repression, the limitations of this pooled approach become clear. The two models offer conflicting interpretations. The OLS model suggests a statistically significant backfire effect from local fatalities ($\beta=0.157$, p = 0.001), while the NB model finds this same relationship to be insignificant (IRR = 1.05, p = 0.283) and instead indicates a significant deterrent effect from violence elsewhere (IRR = 0.90, p = 0.001). This divergence highlights two critical issues that prevent a confident causal interpretation from these models.

First, there is the challenge of \textbf{multicollinearity}. As the momentum results show, prior mobilization is a powerful predictor. It is also, however, highly correlated with the state's repressive response. When two predictors, in this case prior mobilisation and prior repression, are strongly correlated, it becomes statistically difficult for a pooled model to disentangle their independent effects on the outcome. The model struggles to attribute the variance correctly, which can lead to the kind of unstable coefficients and shifting p-values we observe between our OLS and NB specifications.

Second, any pooled model is vulnerable to \textbf{omitted variable bias}. By treating every division-day as an independent observation, these models cannot control for the vast number of unobserved, stable characteristics that make each division unique. Nor can they account for unobserved national-level shocks that affect all divisions on a given day (e.g., a critical news story). Therefore, to separate the real impact of repression from the noise of momentum and other unobserved factors, we must use a more powerful panel data model.

\subsection{Causal Inference: The Repression Backfire Effect}

The ambiguity of the pooled models necessitates a more robust approach. To isolate the causal effect of repression from the confounding factors identified previously, we now turn to our Two-Way Fixed Effects (TWFE) panel model. The strength of this model lies in its ability to control for unobserved sources of bias. By including fixed effects for each of the eight administrative divisions, we account for all stable characteristics that might make one region more prone to protest than another. Simultaneously, by including fixed effects for each day in our sample, we control for any national-level events or shocks that would affect all divisions at the same time. This powerful design allows us to move beyond simple correlation and estimate the direct, localised effect of a change in repression on a subsequent change in mobilisation.

The model includes a full set of dummy variables for each of the eight administrative divisions (entity fixed effects) and for each day in our sample (time fixed effects). Any predictor from our pooled models that does not vary within both of these dimensions is necessarily dropped from the analysis due to perfect collinearity. Specifically, the \textbf{entity fixed effects} absorb the influence of all stable, time-invariant characteristics of each region. This means variables like `Population`, which are constant for a given division, are rendered redundant. The \textbf{time fixed effects} absorb the influence of all national-level shocks or trends that affect all divisions on a given day. This renders variables like `Days\_since\_start`, `Elsewhere\_fatalities`, and `Elsewhere\_events` redundant, as their values are constant across all divisions at each point in time. This process of ``absorption'' leaves us with only the predictors that vary across both divisions and time: `$LocalFatal_{t-1}$` and `$LocalEvents_{t-1}$'. This powerful design allows us to move beyond simple correlation and estimate the direct, localised effect of a change in local repression on a subsequent change in local mobilisation, net of all confounding regional and national factors.

The results of the TWFE model are presented in Table \ref{tab:twfe_model}. The findings are clear and decisive, providing a powerful answer to the puzzle that the pooled models could not solve.

\begin{table}[h!]
\centering
\caption{Two-Way Fixed Effects Panel Model Predicting Mobilisation}
\label{tab:twfe_model}
\begin{tabular}{lc}
\hline \hline
 & \textbf{(1) TWFE OLS} \\
 & $\sqrt{\text{Events}}$ ($\beta$) \\
\hline
 & \\
$\sqrt{\text{LocalFatal}_{t-1}}$ & 0.2221$^{***}$ \\
 & (0.0489) \\
 & \\
$\sqrt{\text{LocalEvents}_{t-1}}$ & 0.0740 \\
 & (0.0560) \\
 & \\
\hline
Observations & 488 \\
R-squared (Within) & 0.1580 \\
Entity Fixed Effects & Yes \\
Time Fixed Effects & Yes \\
\hline \hline
\multicolumn{2}{l}{\footnotesize Robust standard errors in parentheses.} \\
\multicolumn{2}{l}{\footnotesize $^{*}p<0.1$; $^{**}p<0.05$; $^{***}p<0.01$} \\
\end{tabular}
\end{table}

The primary result from this model is the coefficient on our measure of local repression. We find that the effect of prior-day local fatalities on current-day protest mobilization is positive and highly statistically significant ($\beta = 0.2221$, $p < 0.001$). This provides strong, robust support for our second hypothesis (H2). After stripping out all the confounding noise from stable regional differences and shared national trends, the underlying dynamic becomes clear: local state killings directly and significantly fueled subsequent local protest. This is a clear causal evidence for a repression backfire effect.

Interestingly, the model also reveals a new insight into the role of momentum. The coefficient for prior-day local events, which was overwhelmingly significant in the pooled models, is now statistically insignificant (p = 0.187). This does not mean momentum was unimportant. Rather, it suggests that what appeared to be purely local momentum in the pooled models was largely a reflection of each division's participation in a shared, \textit{national} wave of protest. The time fixed effects have absorbed this national trend, revealing that the purely localized, day-to-day momentum was a less powerful driver than the direct reaction to state violence.

While this model confirms that a \textit{backfire effect} exists, it estimates an average effect across the entire revolutionary period. Our qualitative narrative, however, suggests that the backfire was not a constant, but a dynamic that was \textit{triggered} after a catalytic moral shock. This leads us to our next and most critical question: was this backfire effect present from the start, or was it the result of a fundamental shift in the conflict? To answer this, we now turn to a \textbf{structural break analysis}.

\subsection{The Non-Linear Dynamic: A Catalytic Moral Shock}

While our main fixed effects model confirms an average backfire effect, our qualitative narrative suggests a more dramatic, non-linear story. We hypothesised (H3) that the backfire was not a constant feature of the conflict, but a contingent dynamic \textit{triggered} by a catalytic moral shock. To formally test this, and to pinpoint the timing of this shift, we augment our TWFE model with an interaction term and test two distinct, historically-grounded cutoff dates. Table \ref{tab:structural_break} presents these results side-by-side.

Our primary specification (Column 1) sets the break point on \textbf{July 16th}, the day of the first widely publicised protester death, that of Abu Sayeed. Crucially, the event was captured on video and circulated widely online, creating a powerful and visceral visual of an unarmed student being killed by state forces. These images and videos, more than any abstract casualty report, left a deep scar in the public consciousness, crystallising the state's brutality and taking the national outrage to an uncontrollable level. A key analytical advantage of this date is that it allows us to test for a change in mobilisation dynamics at the precise moment this potent visual evidence entered the public sphere. As a point of comparison, we also test a cutoff on \textbf{July 19th} (Column 2), a date which saw the most deaths in the whole of July, representing the very peak of state repression.

\begin{table}[h!]
\centering
\caption{TWFE Structural Break Models Predicting Mobilization}
\label{tab:structural_break}
\begin{tabular}{lcc}
\hline \hline
 & \textbf{(1)} & \textbf{(2)} \\
 & \multicolumn{2}{c}{$\sqrt{\text{Events}}$ ($\beta$)} \\
\cline{2-3}
 & Cutoff: July 16 & Cutoff: July 19 \\
\hline
 & & \\
$\sqrt{\text{LocalFatal}_{t-1}}$ (Pre-Break Effect) & -0.0720 & 0.0306 \\
 & (0.1229) & (0.1136) \\
 & & \\
$\text{Interaction}_{\text{Fatalities} \times \text{Post-Break}}$ & 0.3278$^{***}$ & 0.2169$^{*}$ \\
 & (0.1258) & (0.1162) \\
 & & \\
$\sqrt{\text{LocalEvents}_{t-1}}$ & 0.0834 & 0.0833 \\
 & (0.0557) & (0.0560) \\
 & & \\
\hline
\textit{Derived Total Effect (Post-Break)} & & \\
Total Effect of $\sqrt{\text{LocalFatal}_{t-1}}$ & 0.2558 & 0.2474 \\
\hline
Observations & 488 & 488 \\
R-squared (Within) & 0.1877 & 0.1788 \\
Entity Fixed Effects & Yes & Yes \\
Time Fixed Effects & Yes & Yes \\
\hline \hline
\multicolumn{3}{l}{\footnotesize Robust standard errors in parentheses.} \\
\multicolumn{3}{l}{\footnotesize  $^{*}p<0.1$; $^{**}p<0.05$; $^{***}p<0.01$} \\
\end{tabular}
\end{table}

The results from these models tell a clear and consistent story. Across both specifications, the coefficient for the main repression variable (the ``Pre-Break Effect'') is statistically insignificant. This confirms that in the early, low-intensity phase of the protests, there was no significant backfire effect. State violence, at this stage, did not systematically lead to increased mobilization.

The crucial finding is the comparison of the interaction terms. In our primary specification with the July 16th cutoff (Column 1), this term is positive, substantially large, and highly statistically significant ($\beta = 0.3278$, p = 0.0095). This provides powerful statistical proof that the relationship between repression and mobilisation was fundamentally and significantly different after the shock. When we move the cutoff to July 19th (Column 2), the interaction term remains positive and substantial ($\beta = 0.2169$), but its statistical significance is reduced by a noticeable margin (p = 0.063). It shows that the \textit{visuals} of Abu Sayeed's death had more impact on mobilization compared to the high \textit{number of deaths} that was seen during the curfew and internet shutdown, when visuals were harder to circulate.

A crucial question for any fixed effects model is how well it explains the data once the major sources of variation have been controlled for. This is measured by the R-squared (Within). Our model for the July 16th structural break yields a value of 0.1877. This tells us that our key variables, local repression and its interaction with the catalytic moral shock, accounts for a remarkable 18.77\% of the day-to-day dynamics of protest within each division. To be able to explain nearly a fifth of the variance in such a chaotic environment, after already removing all baseline regional and temporal effects, points to a strong and meaningful underlying relationship~\cite{OziliRSquare2022}.

Taken together, these results allow us to pinpoint the nature of the catalyst with confidence. The sharpest and most statistically significant shift in protester behaviour was triggered by the initial wave of lethal violence around \textbf{July 16th}, as visualised by the dramatic geographic expansion of protest in Figure \ref{fig:maps}. It was this first shock of lethal force, symbolised by the death of Abu Sayeed, along with its disturbing and brutal visuals circulated as a viral video on Facebook, that unlocked a powerful backfire dynamic that was previously absent. While the subsequent institutional repression was also part of the conflict, it was the initial bloodshed and the emotional impact of its visual representation that appears to have been the primary catalyst, turning state violence from an ineffective deterrent into a potent driver of the revolutionary cascade that followed. The brutality of the lethal forces applied on Abu Sayeed and other protesters \textit{shocked} the public's conscience and awakened their \textit{morality}, leading to an exponential increase in mobilization in the days that followed, underscoring our \textbf{catalytic moral shock} hypothesis. 

\subsection{Dynamic Feedback: Visualising the Action-Reaction Cycle}

Our fixed effects analysis provided strong causal evidence for the backfire effect. To complement this finding and to visualize the day-to-day dynamic interplay between state repression and mobilisation at the national level, we needed a more dynamic tool. For this, we turn to a Vector Autoregression (VAR) model, specified on our national-level time-series data. After ensuring both our `total\_fatalities` and `mobilization\_events` series were stationary by applying a first-difference transformation, we fit the model. An analysis of the information criteria suggested a complex, long-memory process, leading us to select an optimal lag length of 10 for the final specification.

\begin{figure*}[h!]
    \centering
    \fbox{\includegraphics[width=0.6\textwidth]{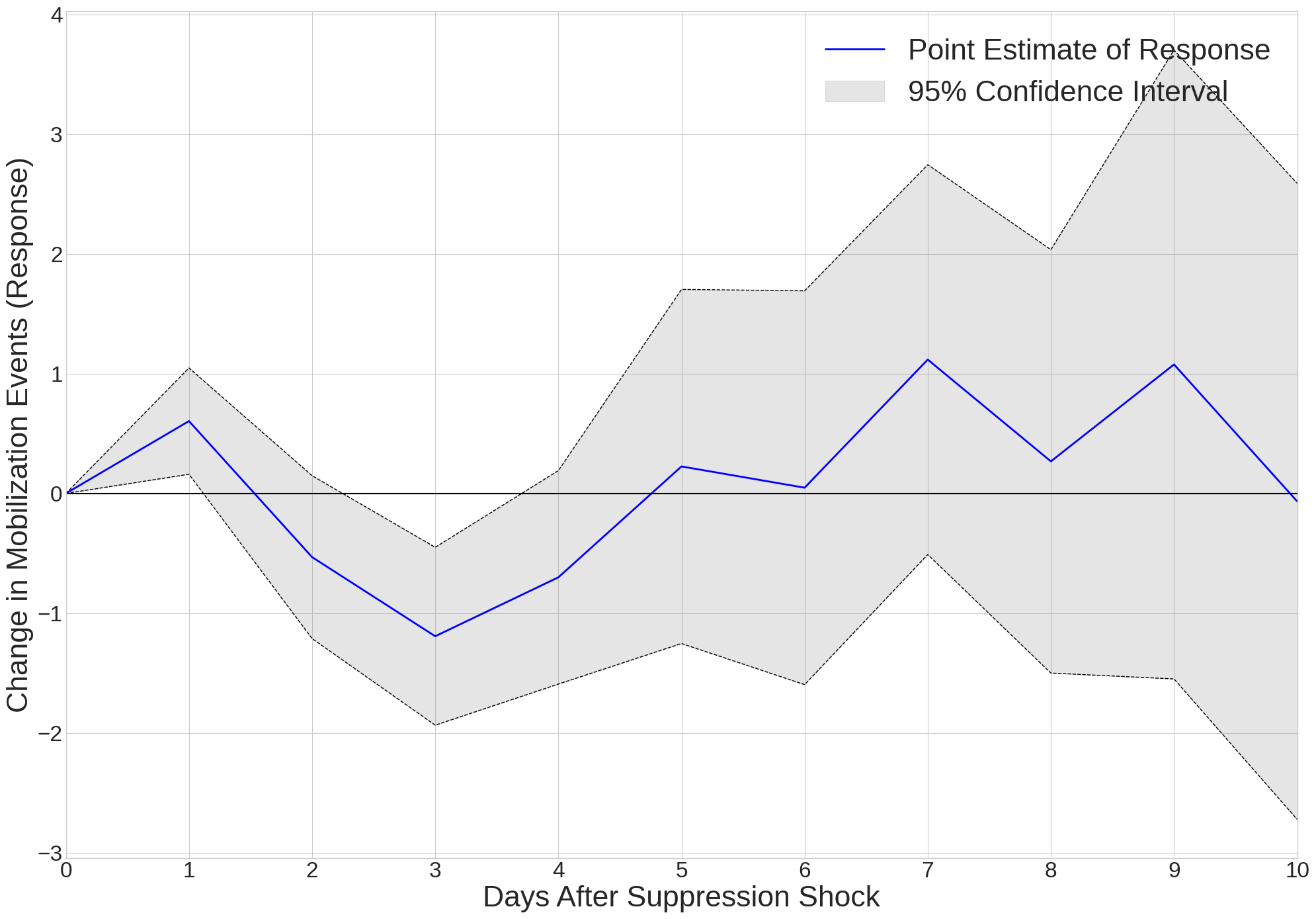}}
    \caption{Impulse Response Function: The Dynamic Response of Mobilization Change to a Suppression Shock. The solid line shows the point estimate of the response; the shaded area represents the 95\% confidence interval.}
    \label{fig:irf}
\end{figure*}

The key result from this model is not a table of coefficients, but a picture: the Impulse Response Function (IRF), shown in Figure \ref{fig:irf}. The IRF provides a powerful, intuitive visualization of the conflict's feedback loop. It answers a simple question: when there is a sudden, unexpected shock of state violence, how do the streets react over the next ten days?

The story told by the IRF is one of immediate and powerful backlash. As seen in Figure \ref{fig:irf}, an unexpected increase (a \textit{shock}) in fatalities is met with a positive and statistically significant increase in mobilisation on the \textbf{very next day} (Day 1). The confidence interval at this first lag is entirely above the zero line, confirming a rapid and potent backfire. This initial surge of outrage is followed by a more complex dynamic. The point estimate suggests a potential dampening effect around Day 3. It points to a potential cyclical pattern of protest and pause before the system returns to its baseline.

This result is crucial for two reasons. First, it corroborates our main causal finding from an entirely different methodological angle, greatly strengthening our confidence in our second hypothesis (Backfire effect). Second, it provides a vivid visualization of the \textit{immediacy} of this effect. The feedback loop between state violence and popular resistance was incredibly tight. It shows that the regime's repressive actions were not just counterproductive in the long run; they translated into a powerful surge of defiance on the streets in less than 24 hours.

\subsection{Predictive Modelling: Signals of Escalation}

In our final analytical step, we move from causal inference to a more exploratory, predictive lens. The goal here is to answer a different question: setting aside the constraints of a formal causal model, what were the most powerful raw signals for predicting the next wave of protest? This is a task for which modern machine learning models are exceptionally well-suited. While more advanced deep learning architectures are often used for time-series forecasting~\cite{QaziEventForecast2024},~\cite{SaitoTwitterSentiment2022}, they are notoriously data-hungry and at high risk of overfitting on a short, explosive event like the July Revolution. We therefore turn to two powerful ensemble models, XGBoost and Random Forest, which are state-of-the-art for tabular data of this scale. Our final feature set was manually curated to mitigate the severe multicollinearity present in the initial data. It ensures the interpretability of our feature importance rankings. This approach proved superior not only for interpretation but also for prediction, boosting the out-of-sample R-squared from \textbf{0.606} to a final \textbf{0.646}.

To rigorously test the predictive power of the historical data, we employed a walk-forward cross-validation for both models. The results, summarised in Table \ref{tab:ml_metrics}, were impressive. Both models demonstrated a strong ability to forecast the next day's protest activity, confirming that the revolution, while seemingly chaotic, was driven by highly structured and predictable patterns. The XGBoost model proved to be the superior predictor in terms of pure accuracy, explaining nearly 65\% of the variance in a true out-of-sample test.

\begin{table}[h!]
\centering
\caption{Predictive Performance of Machine Learning Models}
\label{tab:ml_metrics}
\begin{tabular}{lcc}
\hline \hline
 & \textbf{(1) XGBoost} & \textbf{(2) Random Forest} \\
\hline
R-squared (Out-of-Sample) & 0.646 & 0.520 \\
MAE & 8.266 & 8.402 \\
RMSE & 13.664 & 15.918 \\
\hline
Most Important Feature & `Excessive force` & `Excessive force` \\
(Aggregated Importance) & (0.915) & (0.528) \\
\hline \hline
\multicolumn{3}{l}{\footnotesize Metrics from a 2-day lag walk-forward cross-validation.} \\
\end{tabular}
\end{table}

Beyond pure prediction, the true value of these models lies in their ability to identify the most important predictive signals. While both models agree on the most critical feature, their different algorithms provide complementary insights. XGBoost, a boosting model that builds trees sequentially, tends to latch onto and exploit the single strongest predictor. As shown in Table \ref{tab:ml_metrics}, it attributes over 91\% of its predictive power to a single, \textbf{dominant predictor}: the recent history of `Excessive force against protesters`. This particular sub-event category specifically captures the events where "authorities or other organized groups use violent force to disperse or contain a protest", according to the ACLED codebook~\cite{acled_source}, rather than just the death toll. Our model's overwhelming reliance on this feature, therefore, provides strong quantitative evidence that the primary catalyst for the revolution's escalation was not the abstract knowledge of casualties, but the overwhelming reaction to the visible act of state brutality.

\begin{figure*}[h!]
    \centering
    \fbox{\includegraphics[width=0.65\textwidth]{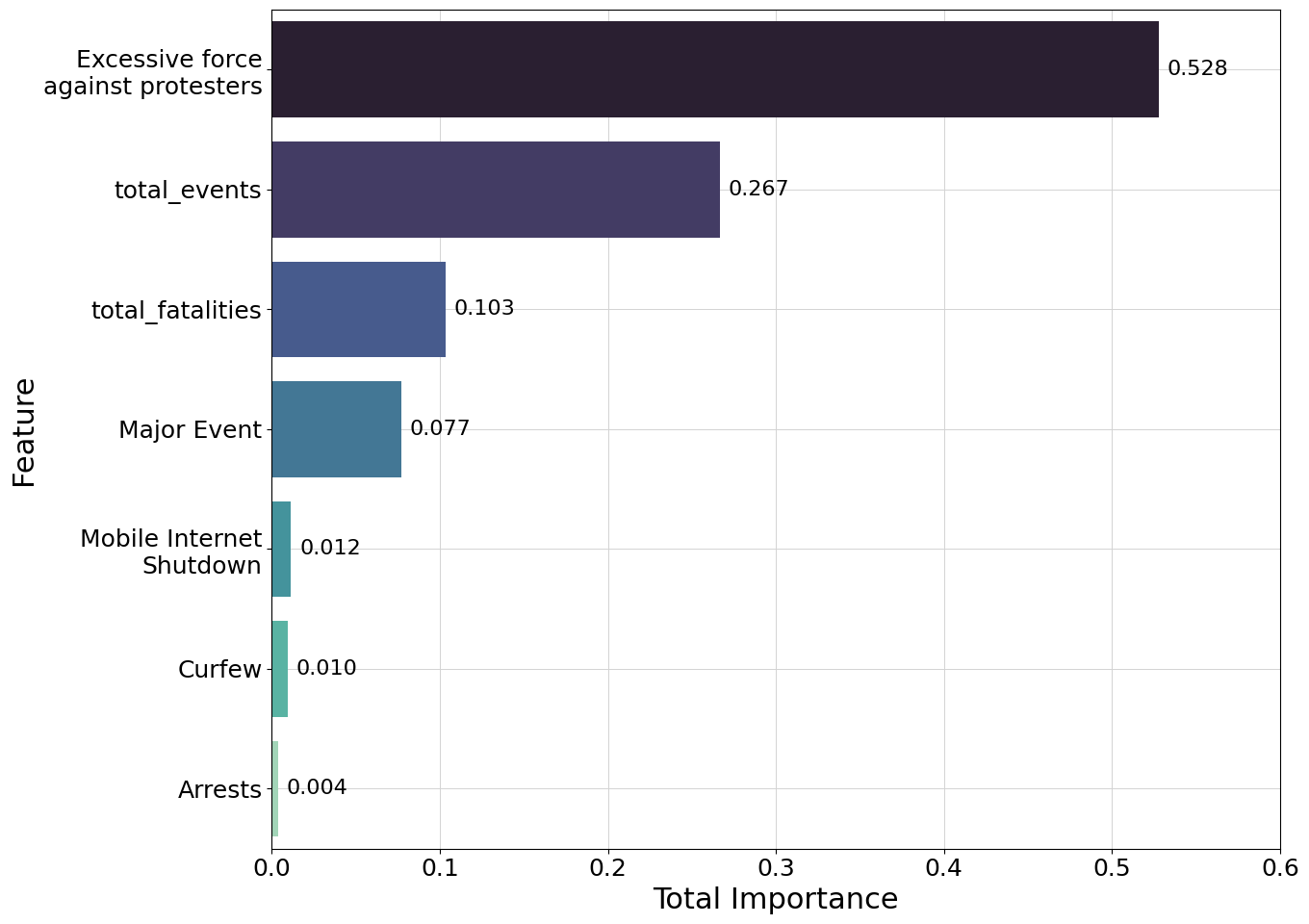}} 
    \caption{Aggregated Feature Importance for Predicting Total Events (Random Forest). Importance scores for each base feature are aggregated across their 1-day and 2-day lags.}
    \label{fig:rf_importance}
\end{figure*}

The Random Forest model, a bagging algorithm that builds trees independently on random subsets of features, provides a more nuanced and distributed perspective. It corroborates that \textbf{`Excessive force against protesters`} is the single most important predictor, but it also reveals a ``chorus'' of other powerful signals. Figure \ref{fig:rf_importance} displays the importance of the lagged features from the Random Forest model. After the spectacle of repression, the next most important predictors are \textbf{protest momentum} (`total\_events`), the direct impact of \textbf{lethal force} (`total\_fatalities`), and the occurrence of a \textbf{Major Event}. These important features align perfectly with our three hypotheses: \textbf{H1} (Momentum), \textbf{H2} (Repression Backfire), and \textbf{H3} (Catalytic Shock/Event acting as a Trigger).

Taken together, our causal inference models and these predictive models complement one another strongly, showing that while the revolution was a complex phenomenon, its escalations were best predicted by the recent history of visible state brutality. This finding strongly suggests that the primary catalyst was not just the abstract knowledge of casualties, but the visceral, visual spectacle of repression. This \textbf{dominant predictor}, backed by the chorus of momentum, lethal force, and major events, provided the fuel for the revolutionary fire.

\section{Conclusion}
This paper set out to dissect the repression-mobilisation dynamic at the heart of Bangladesh's historic 2024 July Revolution. To do so, we developed a mixed-methods framework, using a qualitative historical narrative to derive three core hypotheses regarding momentum, backfire, and non-linearity, which we then tested from multiple angles using a layered quantitative approach. Our findings, validated across a suite of econometric and machine learning models, provide a robust and nuanced account of how a student protest transformed into a successful, nationwide uprising.

Our analysis confirms all three hypotheses. We found that the movement was sustained by its own powerful momentum, but that its explosive growth was driven by a \textbf{contingent backfire} effect. Our Two-Way Fixed Effects model provided strong causal evidence for this backfire at the local level, and our structural break analysis proved this dynamic was non-linear, demonstrating that it was triggered by the moral shock of the first lethal violence and its viral spectacles around July 16th. These findings were complemented by our powerful predictive models too.

The story of the July Revolution is therefore one where the catalytic moral shock stemming from viral, visual evidence of state brutality became the primary driving force behind a revolutionary cascade. This dynamic is not unique to Bangladesh. It echoes the dynamics of other successful modern uprisings in the region, from the recent student movement that toppled the regime in Nepal in 2025 to the 2022 Aragalaya in Sri Lanka, where the state's visible brutality was the final straw that united the public against it. Our findings suggest that in an interconnected, digital age, the ability of protesters to document and rapidly circulate evidence of repression can fundamentally alter the traditional dynamics of power.

Methodologically, our work offers a robust and replicable framework for analysing modern short-duration, high-intensity movements. By integrating causal models (Fixed Effects), dynamic models (VAR), and predictive models (XGBoost, RF), we demonstrate how researchers can rigorously test and validate their hypotheses from multiple perspectives. While such analysis is necessarily a simplification of a complex human event, it provides a powerful, data-driven foundation for understanding the mechanisms of modern civil resistance and the conditions under which authoritarian control can so spectacularly fail.

\section*{Data and Code Availability}

The complete source codes and modified dataset for our analyses are publicly available on GitHub at: \url{https://github.com/Saiful185/July-Revolution-Analysis}.

\section*{Acknowledgement}

The authors wish to express their sincere gratitude to Dr. Michael Biggs and Dr. Mathis Ebbinghaus of the Department of Sociology, University of Oxford, whose constructive feedback and methodological guidance were invaluable in refining the research design and analytical approach of this paper.

 

\newpage

\section*{Author Biographies}
\begin{IEEEbiography}[{\includegraphics[width=1in,height=1.25in,clip,keepaspectratio]{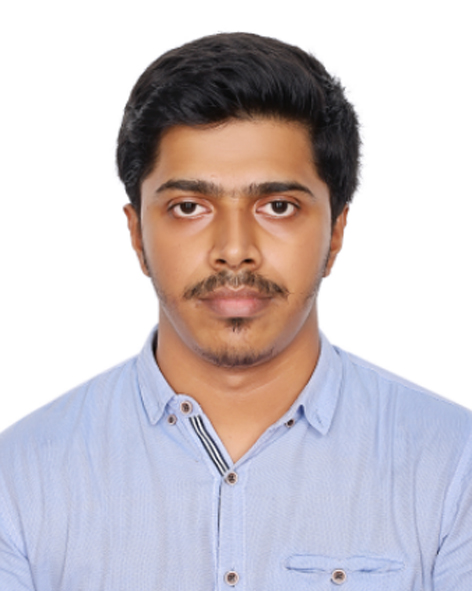}}]{Md. Saiful Bari Siddiqui}
(Graduate Student Member, IEEE) received the B.Sc. degree in Electrical and Electronic Engineering from the Bangladesh University of Engineering and Technology in 2021 and the M.Sc. degree in Computer Science and Engineering from BRAC University, Dhaka, Bangladesh, in 2024. He is currently a Senior Lecturer in the Department of Computer Science and Engineering at BRAC University, where he is involved in teaching, mentoring, and research. He has previously served as a Research Assistant at the Centre for Cognitive and Data Sciences (CCDS), Independent University, Bangladesh, contributing to interdisciplinary projects at the intersection of cognitive science and data analytics. His research interests include machine learning, deep learning, computational social science, and applications of artificial intelligence in medical imaging and real-world datasets. He has also been involved in collaborative research projects supported by organizations like the Institute for Advanced Research (IAR), United International University.
\end{IEEEbiography}

\begin{IEEEbiography}[{\includegraphics[width=1in,height=1.25in,clip,keepaspectratio]{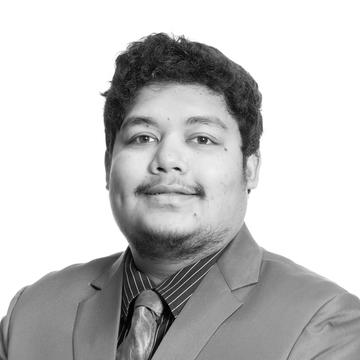}}]{Anupam Debashis Roy}
received the B.A. degree in Interdisciplinary Studies from Howard University, USA and the M.Sc. degree in Political Sociology from London School of Economics, London, UK. He is currently a Ph.D. candidate in the Department of Sociology at the University of Oxford, London, UK. His work has appeared in South Asia Research, NYU Undergraduate Law Review, Himal Southasian, The Daily Star, Suddhashar, Frontline-The Hindu, Dhaka Tribune, Prothom Alo, Ittefaq etc. His 2020 book Not All Spring Ends Winter on Bangladeshi social movements received acclamation from readers and critics and has been used as a course textbook at universities. Anupam has conducted research on the 2013 Shahbag Movement and the 2018 Road Safety Movement in Bangladesh in his undergraduate and master’s theses respectively and is currently analysing the 2024 July Uprising. His DPhil thesis will explore the similarities and differences among these movements to ascertain why some movements were more “successful” than others. His tentative ideas about social movements include the theories of cumulative learning and strategic malleability of movement tactics. He aspires to use empirical data from Bangladesh to extrapolate novel theories on social movement and make an original contribution to existing knowledge.
\end{IEEEbiography}

\vfill

\end{document}